\documentclass[aps, twocolumn,citeautoscript]{revtex4-2}
\setcitestyle{super}
\usepackage{amsmath}
\usepackage{braket}
\usepackage{array}
\usepackage{graphicx}
\usepackage{float}
\usepackage{booktabs}
\usepackage{hyperref}
\usepackage[utf8]{inputenc}
\usepackage{soul,xcolor}
\setstcolor{red}
\begin{document}
\title{New Local Explorations of the Unitary Coupled Cluster Energy Landscape}
\author{Harper R. Grimsley}
\affiliation{Chemistry Department, Virginia Tech}
\author{Nicholas J. Mayhall}
\affiliation{Chemistry Department, Virginia Tech}
\email{nmayhall@vt.edu}
\date{\today}
\begin{abstract}
The recent quantum information boom has effected a resurgence of interest in unitary coupled cluster (UCC) theory.  Our group's interest in local energy landscapes of unitary ans{\"a}tze prompted us to investigate the \textit{classical} approach of truncating the Taylor series expansion (instead of a perturbative expansion) of UCCSD energy at second-order.  This amounts to an approach where electron correlation energy is estimated by taking a single Newton-Raphson step from Hartree-Fock toward UCCSD. Such an approach has been explored previously, but the accuracy was not extensively studied. In this paper, we investigate the performance and observe similar pathologies to linearized coupled cluster with singles and doubles.  We introduce the use of derivatives of order three or greater to help partially recover the variational lower bound of true UCCSD, restricting these derivatives to those of the ``unmixed'' category in order to simplify the model.  By testing the approach on several potential energy surfaces and reaction energies, we find this ``diagonal'' approximation to higher order terms to be effective at reducing sensitivity near singularities for strongly correlated regimes, while not significantly diminishing the accuracy of weakly correlated systems. 

\end{abstract}
\maketitle
\section*{Introduction}
Unitary coupled cluster (UCC) has intrigued chemists for the past four decades as a variational form of the already-powerful coupled cluster theory. \cite{kutzelnigg_w_52_1977, bartlett_alternative_1989, kutzelnigg_error_1991, taube_rethinking_2009, harsha_difference_2018, chen_quantum-inspired_2021}  It has garnered significant attention in recent years due to its use in variational quantum eigensolvers (VQEs), hybrid quantum-classical algorithms.\cite{peruzzo_variational_2014, hempel_quantum_2018, nam_ground-state_2020}  Quantum gates must correspond to physical evolutions, making a unitary parametrization of the wavefunction a natural choice.  Various approximations and modifications of UCC have seen use, including more general ans{\"a}tze based on UCC-like generators\cite{hatano_finding_2005, mcclean_theory_2016, romero_strategies_2018, ryabinkin_qubit_2018,lee_generalized_2019,ryabinkin_iterative_2020, grimsley_is_2020}.  A comprehensive review of this topic has recently been prepared by Anand et. al.\cite{anand_quantum_2022}  Dynamical ans\"atze such as the Adaptive, Derivative-Assembled, Pseudo-Trotterized VQE (ADAPT-VQE) method\cite{grimsley_adaptive_2019,tang_qubit-adapt-vqe_2021} build a sequential, UCC-like ansatz one anti-Hermitian operator at a time, the order of which is determined by the energy gradient associated with that operator pair's introduction to the ansatz.  ADAPT-VQE gives a quasi-optimal operator ordering that is informed by the problem Hamiltonian, but only includes information about the first energy derivative with respect to the energy parameters.  

While UCC has provided a useful framework for defining state preparation circuits in VQE, as a classical approach it also has several desirable attributes: 
\begin{enumerate}
    \item Variationality
    \item Size-extensivity \cite{taube_new_2006}
    \item Satisfaction of the generalized Hellmann-Feynman theorem\cite{bartlett_alternative_1989}
\end{enumerate}
Regardless of its various attractive features, computing the UCC energy requires evaluating an infinite series of commutators, even if the excitation rank of $\hat{T}$ is restricted to singles and doubles (UCCSD).  
The UCCSD energy landscape, while complicated, is a smooth functional of the UCCSD wavefunction $t$-amplitudes.  
If we assume that we are sufficiently close to the energy minimum that the landscape is locally convex \textcolor{black}{and that a finite solution exists which satisfies the stationary conditions}, we can make a reasonable guess at the optimal values of the amplitudes by using a quadratic Taylor approximation to the energy functional in the $t$-amplitudes.  
In general, we would need a higher-order Taylor expansion to fully understand the landscape, and recover the nice properties of UCC.  
One way we can go beyond the quadratic approximation without much difficulty is to partially explore the cubic and higher characters of the landscape.  
Rather than including all higher order derivative tensors between, we consider only diagonal partial derivatives, e.g.,  $\frac{\partial^k \epsilon}{\left(\partial t_{ij}^{ab}\right)^k}$ beyond second order.  
Such an approximation is based on the assumption that high-order mixed derivatives are not important, a purely geometric idea, rather than one based on, e.g., perturbation theory.  
We will explore this idea more concretely in the theory section.

\section*{Theory}
We will begin this section by introducing the full UCCSD functional, and proceed to its Taylor series truncation at second order in the $t$-amplitudes.  From there, we will introduce our ``diagonal'' correction strategy, and give explicit equations for the third- and infinite-order situations.  We will finally compare the third-order case to existing coupled electron pair approximation (CEPA) methods.

The unitary coupled cluster ansatz is given by Eq. \ref{ucc}\cite{kutzelnigg_w_52_1977}.
\begin{align}\label{ucc}
\ket{\Psi_{UCC}} = e^{\left(\hat{T}-\hat{T}^{\dagger}\right)}\ket{0}
\end{align}
where $\hat{T}$ is defined in the usual way, as in Eq. \ref{t}.
\begin{align}\label{t}
\hat{T} &= \hat{T}_1 + \hat{T}_2 + \dots\\
&= \sum_{ia}t_a^i\hat{a}_i^a + \sum_{\substack{i<j\\a<b}}t_{ab}^{ij}\hat{a}_{ij}^{ab}+\dots \nonumber 
\end{align}
We will define $\ket{0}$ to be a normalized, single-determinant reference.  We do not initially assume a Brillouin condition, 
and will explore the consequences of orbital choice later in the text.   Because $\hat{T}-\hat{T}^{\dagger}$ is anti-Hermitian, its exponential is unitary, and the quantity $\epsilon_{UCC}[\mathbf{t}]$ in Eq. \ref{eps} is bounded below by the lowest eigenvalue of $\hat{H}$ for any value of $\mathbf{t}$, 
\begin{equation}\label{eps}
\epsilon_{UCC}[\mathbf{t}] = \braket{\Psi_{UCC}|\hat{H}|\Psi_{UCC}},
\end{equation}
where $\mathbf{t}$ is the vector of $t$ amplitudes.
While $\epsilon_{UCC}$ is variational, symmetric, size-extensive, and satisfies the generalized Hellmann-Feynman theorem, it is classically intractable to evaluate.  
Unlike the traditional (non-unitary) coupled cluster energy, the Baker-Campbell-Hausdorff (BCH) expansion of $\epsilon_{UCC}$ will never terminate, regardless of truncation of $\hat{T}$.  
Various artificial truncation schemes for $\epsilon_{UCC}$ exist, including truncation based on perturbation order\cite{bartlett_alternative_1989} and truncation based on commutator order.\cite{kutzelnigg_w_52_1977}  
We will restrict our focus to the latter, which has a geometric interpretation and has been examined in far less detail. 
To denote order ($n$) of Taylor series trunctation, we prepend an ``O$n$-'' to the typical UCCSD.
For example, restricting $\hat{T}$ to singles and doubles,
and terminating the Taylor series after second order,
we obtain Eq. \ref{functional}, which in this notation is given as O2-UCCSD.
\onecolumngrid
\begin{align}\label{functional}
    \epsilon_{O2-UCCSD} &= E_0 + \sum_{ia}\left.\frac{\partial \epsilon_{UCCSD}}{\partial t_a^i}\right|_\mathbf{0} t_a^i + \sum_{\substack{i<j\\a<b}}\left.\frac{\partial \epsilon_{UCCSD}}{\partial t_{ab}^{ij}}\right|_\mathbf{0} t_{ab}^{ij}\\
 &+\frac{1}{2}\sum_{ijab}\left.\frac{\partial^2\epsilon_{UCCSD}}{\partial t_a^i \partial t_b^j}\right|_\mathbf{0} t_a^i t_b^j +\frac{1}{2}\sum_{\substack{i<j,k<l\\a<b,c<d}}\left.\frac{\partial^2\epsilon_{UCCSD}}{\partial t_{ab}^{ij}\partial t_{cd}^{kl}}\right|_\mathbf{0} t_{ab}^{ij} t_{cd}^{kl} +\sum_{\substack{i<j\\a<b\\kc}}\left.\frac{\partial^2\epsilon_{UCCSD}}{\partial t_{ab}^{ij}\partial t_c^k}\right|_\mathbf{0}t_{ab}^{ij}t_c^k \nonumber \\
&= E_0 + 2\braket{0|\hat{H}_N\hat{T}|0} + \braket{0|\hat{T}^\dagger\hat{H}_N\hat{T}|0} + \braket{0|\hat{H}_N\hat{T}_1^2|0} - \braket{0|\hat{H}_N\hat{T}_1^\dagger \hat{T}_2|0}\nonumber
\end{align}
\twocolumngrid
(We use  the standard definitions of $\hat{F}_N$, $\hat{V}_N$, and $\hat{H}_N$, as used in the coupled cluster review of Crawford and Schaefer.\cite{crawford_introduction_2007})  Apart from the last two terms, $\epsilon_{O2-UCCSD}$ is simply the LCCSD Lagrangian.\cite{pulay_gradients_1983}  Similarly, $\epsilon_{O2-UCCD}$ is precisely the LCCD Lagrangian.  Minimizing $\epsilon_{O2-UCCSD}$ amounts to solving a linear set of stationary equations, given by equations \ref{singles} and \ref{doubles}.
\begin{align}
	\braket{\phi_i^a|\hat{H}_N\hat{T}_1+\hat{T}_1^\dagger \hat{V}_N+\frac{1}{2}\hat{F}_N\hat{T}_2|0} &= -f_i^a \label{singles}\\
	\braket{\phi_{ij}^{ab}|\hat{H}_N\hat{T}_2+\frac{1}{2}\hat{F}_N\hat{T}_1|0} &= -\braket{ij||ab} \label{doubles}
\end{align}
For non-Hartree-Fock orbitals, the lack of a Brillouin condition,
leads to a breakdown of size-extensivity when differentiating the
$t_{ij}^{ab}t_{i}^af_{jb}$ 
term with respect to the doubles amplitudes,
resulting in a disconnected contribution.\cite{szalay_alternative_1995}  
This term ultimately persists due to an incomplete cancellation between diagrams in the $\braket{0|\hat{T}^\dagger \hat{H}_N\hat{T}|0}$
and
$\braket{0|\hat{H}_N\hat{T}_1^\dagger\hat{T}_2|0}$
terms.
While this might not be expected to deteriorate performance too significantly, as the magnitude of the size-inextensivity is determined only by the occupied-virtual block of the \textcolor{black}{Fock} matrix (i.e., the distance from an optimal set of orbitals), 
it is, in fact, possible to reformulate the problem slightly to recover exact size-extensivity, for both HF, and non-HF orbitals. 
If we, instead, begin from the partially Trotterized (excitation rank separated) energy in Eq. \ref{trotter}, 
\begin{align}\label{trotter}
	\epsilon_{tUCCSD} = \braket{0|e^{-\hat{K}_2}e^{-\hat{K}_1}\hat{H}e^{\hat{K}_1}e^{\hat{K}_2}|0},
\end{align}
where $\hat{K}_i = \hat{T}_i - \hat{T}_i^\dagger$, we will obtain the following second-order Taylor series approximation,
\begin{align}
    \epsilon_{tO2-UCCSD} =& E_0 + 2\braket{0|\hat{H}_N\hat{T}|0} + \braket{0|\hat{T}^\dagger\hat{H}_N\hat{T}|0}\nonumber\\ 
    &+ \braket{0|\hat{H}_N\hat{T}_1^2|0} - 2\braket{0|\hat{H}_N\hat{T}_1^\dagger \hat{T}_2|0}.
\end{align}
This additional factor of 2, provides full cancellation of the disconnected terms in the gradient expression. 

This is interesting from three perspectives:
(i) any single-determinantal reference state leads to a size-extensive method which becomes equivalent to the first approach when HF orbitals are used,
(ii) the disentangled form suggests the opportunity to develop a proper exact singles approach, since the unitary formalism allows implementation by a simple orbital rotation, 
which will be considered in follow-up work, 
and (iii) a doubles-then-singles ordering of of excitations is consistent with what we find to be accurate in both our ADAPT-VQE algorithm and with our previous direct study on Trotter ordering.\cite{grimsley_is_2020}  
The resulting stationary conditions are given as,
\begin{align}
	\braket{\phi_i^a|\hat{H}_N\hat{T}_1+\hat{T}_1^\dagger \hat{V}_N|0}&= -f_i^a \label{stat1}\\
	\braket{\phi_{ij}^{ab}|\hat{H}_N\hat{T}_2|0} &= -\braket{ij||ab}. \label{stat2}
\end{align}
  At the time of writing, another group pointed out that one can perform a Newton step toward tUCCSD instead of UCCSD.\cite{chen_low-depth_2022}  However, they do not make any argument for a specific Trotter ordering, or take interest in the size-extensivity of the solution.  

 One key advantage of the Taylor-truncated UCCSD approaches compared to CEPA approaches is that Taylor-truncated approaches are systematically improvable.  Inclusion of triple or higher excitations is obvious, if expensive, and we can work toward recovering the variational character of UCCSD by including higher ordered terms in the Taylor series.
We note that an analogous systematic improvability also exists within the perturbation-trunctated UCC schemes.

\textcolor{black}{O2-UCCSD has another issue, unrelated to size-extensivity.  Linearized CC approximations have long been known to blow up due to singularities stemming from quasi-degeneracies.}\cite{jankowski_applicability_1980} 
\textcolor{black}{This behavior is observed even in multireference formulations, though inclusion of non-linear terms avoids this issue.} \cite{jankowski_applicability_1992} \textcolor{black}{As we will report, the same problem plagues O2-UCCSD.  One of the most successful strategies for avoiding singularities historically has been the split-amplitude ``almost linear CC'' family of approaches, where some of the t-amplitudes are described by a large fixed part and a small variable component.}\cite{jankowski_approximate_1996,jankowski_approximate_1998}
The {aforementioned} singularity problem \textcolor{black}{can alternatively} be addressed by expanding $\epsilon_{UCCSD}$ to third order in the $t$-amplitudes, but such an expansion would significantly increase the complexity of the resulting expressions (albeit with no net increase in the asymptotic scaling).  
However, if we assume that the third derivative tensor is diagonally-dominant, we can instead approximate these higher order terms with essentially no additional cost by including only the diagonal (or \textit{unmixed}) third derivatives in the Taylor series, as in Eq. \ref{third}.
\onecolumngrid
\begin{align}\label{third}
	\epsilon_{O2D3-UCCSD} &= \epsilon_{O2-UCCSD} + \frac{1}{6}\sum_{ia}\left.\frac{\partial^3 \epsilon_{UCCSD}}{\left(\partial t_a^i\right)^3}\right|_\mathbf{0}\left(t_a^i\right)^3 + \frac{1}{6}\sum_{\substack{i<j\\a<b}}\left.\frac{\partial^3 \epsilon_{UCCSD}}{\left(\partial t_{ab}^{ij}\right)^3}\right|_\mathbf{0}\left(t_{ab}^{ij}\right)^3\\
	&=\epsilon_{O2-UCCSD} - \frac{4}{3}\sum_{ia}f_i^a\left(t_a^i\right)^3 - \frac{4}{3}\sum_{\substack{i<j\\a<b}}\braket{ij||ab}\left(t_{ab}^{ij}\right)^3 \nonumber
\end{align}
\twocolumngrid
Differentiating Eq. \ref{third} introduces non-linear, but diagonal terms.  In practice, we minimize $\epsilon_{O2D3-UCCSD}$ directly to find the O2D3-UCCSD energy, but this could also be achieved by solving a series of ``shifted'' linear equations if desired. 
In general, including diagonal derivatives of higher order will only introduce new 4-index contractions, which are negligible in an $O(N^6)$ algorithm.

We will advocate, as an improved approach, the inclusion of diagonal derivatives to \textit{infinite} order, O2D$\infty$-UCCSD.  This energy is given in Eq. \ref{inf}.  Einstein notation is used for the orbital indices to improve readability.
\onecolumngrid
\begin{align}\label{inf}
\epsilon_{O2D\infty-UCCSD} &= \epsilon_{O2-UCCSD}+\sum_{k=3}^{\infty} \frac{1}{k!}\left(\left.\frac{\partial^k\epsilon_{UCCSD}}{\left(\partial t_a^i\right)^k}\right|_\mathbf{0}\left(t_a^i\right)^k+\frac{1}{4} \left.\frac{\partial^k\epsilon_{UCCSD}}{\left(\partial t_{ab}^{ij}\right)^k} \right|_\mathbf{0}\left(t_{ab}^{ij}\right)^k\right)
\\&= \textcolor{black}{E_0}+\frac{1}{2}\braket{\phi_0|[[\hat{H},\hat{K}],\hat{K}]|\phi_0}\nonumber +\sin(2t_a^i)f_i^a + \frac{1}{4}\sin(2t_{ab}^{ij})\braket{ij||ab}\nonumber\\
&\nonumber +\left(\sin^2(t_a^i)-\left(t_a^i\right)^2\right)\braket{\phi_i^a|\hat{H}_N|\phi_i^a} +\frac{1}{4}\left(\sin^2(t_{ab}^{ij})-\left(t_{ab}^{ij}\right)^2\right)\braket{\phi_{ij}^{ab}|\hat{H}_N|\phi_{ij}^{ab}} \nonumber 
\end{align}
\twocolumngrid
A derivation of Eq. \ref{inf} is given in appendix \ref{inf_der}. Our diagonal energies bear a natural resemblance to the third- and infinite-ordered two-electron UCC energies computed by Kutzelnigg.\cite{kutzelnigg_error_1991} \textcolor{black}{We note in passing that the diagonal corrections to the O2-UCCSD energy functional give a method which is no longer invariant to occupied-occupied and virtual-virtual orbital rotations.}

To recapitulate, the O2(D2)-UCCSD method was previously described by Kutzelnigg\cite{kutzelnigg_w_52_1977} and has already been implemented for multiple reference determinants by Simons and Hoffmann.\cite{hoffmann_unitary_1988}  This method corresponds to truncating the UCCSD functional at second order in the $t$-amplitudes, then minimizing the truncated functional.  Our O2D3- and O2D$\infty$-UCCSD variants include the unmixed or ``diagonal'' derivatives to third and infinite order, respectively.  (E.g., O2D3-UCCSD approximates the third derivative ``jerk'' \textcolor{black}{tensor} {matrix} by its diagonal.)  The deletion of extensivity-violating terms is not new, but our rationalization based on ansatz Trotterization is, and we denote the use of these deletions as O2DX-tUCCSD.

As a final note on the theory involved in these methods, we point out that the O2D3-UCCSD functional gives similar amplitude equations to performing a conventional CEPA derivation, and treating only the EPV terms in which \textit{every} index, occupied and virtual, is exclusion principle-violating.  (The authors recommend the review\cite{wennmohs_comparative_2008} by Wennmohs and Neese of the EPV-based CEPA derivations.  See also appendix \ref{cepa} which follows in their footsteps.)  Our methods might be viewed as maximally simple approaches that still give unique, determinant-tailored shifts to individual excitations.  This characterization is consistent with their improved capacity to break single bonds relative to LCCSD.\cite{malrieu_ability_2010}  Our method might be compared to a minor complication of LCCSD\cite{purvis_reduced_1981}, a minor simplification of CEPA(2)/CEPA(3),\cite{ahlrichs_many_1979} or a dramatic simplification of SC$^2$CISD.\cite{daudey_sizeconsistent_1998} 
\section*{Results and Discussion}
We consider three potential energy surfaces used by Malrieu et. al.\cite{malrieu_ability_2010}  to demonstrate the single bond-breaking ability of CEPA(3):
\begin{enumerate}
\item Dissociation of hydrogen fluoride
\item Dissociation of a single C-H bond in methane
\item Torsion of ethylene
\end{enumerate}
We used optimized geometries from B3LYP\cite{becke_densityfunctional_1993}/6-31G$^*$\cite{dill_selfconsistent_1975, ditchfield_selfconsistent_1971, hariharan_influence_1973, hehre_selfconsistent_1972} calculations to determine the positions of atoms which were held static in each curve.  All O2-UCCSD energies, as well as LCCSD energies, were computed with a custom software package developed in-house, available at \url{https://github.com/hrgrimsl/taylor_ucc}.  SCF, CCSD, CCSD(T), DFT, and FCI calculations were performed with PySCF.\cite{sun_pyscf_2018}  CCSDT energies were obtained with the MRCC code of K{\'a}llay et. al. \cite{kallay_higher_2001,kallay_mrcc_2020,noauthor_mrcc_nodate} via a Psi4 interface.\cite{parrish_psi4_2017}
\begin{center}
\begin{figure}
\includegraphics[width=.45\textwidth]{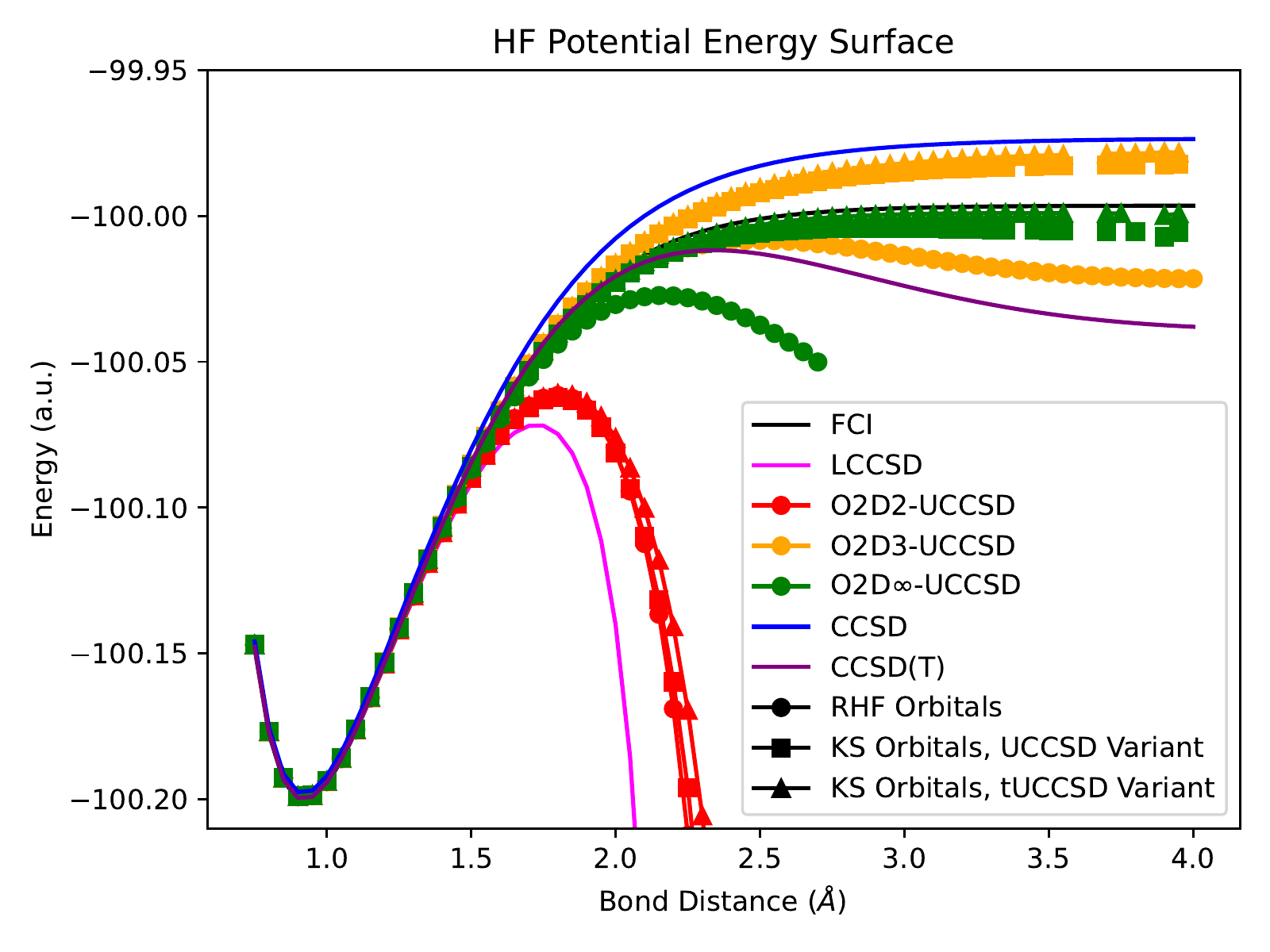}
\caption{Dissociation of HF in the 6-31G$^{**}$\cite{ditchfield_selfconsistent_1971, hariharan_influence_1973, hehre_selfconsistent_1972} basis.}
\label{hf}
\end{figure}

\begin{figure}
\includegraphics[width=.45\textwidth]{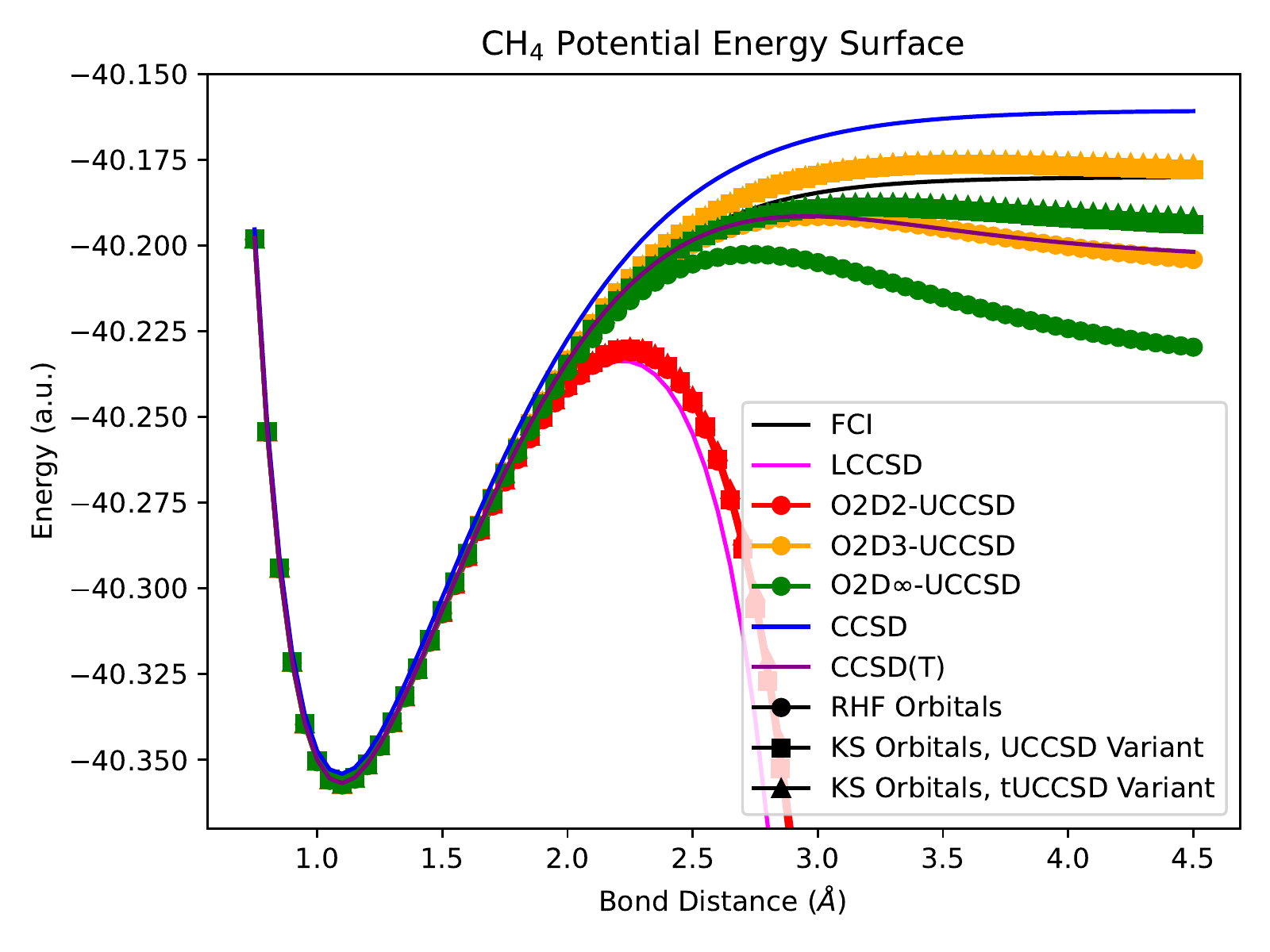}
\caption{Dissociation of a single C-H Bond in methane in the 6-31G$^*$ basis.}
\label{ch4}
\end{figure}
\begin{figure}
\includegraphics[width=.45\textwidth]{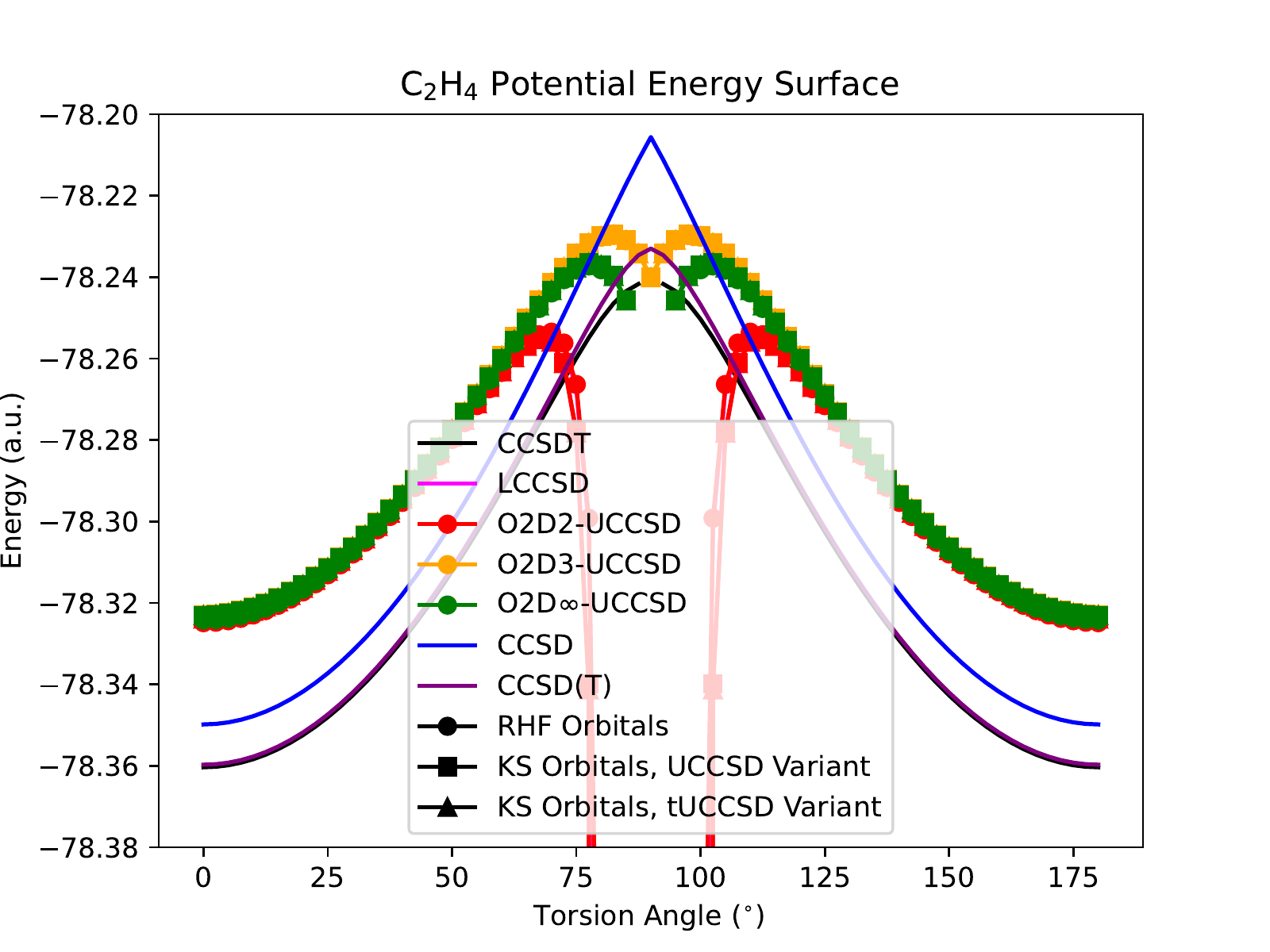}
\caption{Rotation of one CH$_2$ group about the C-C bonding axis in ethylene in the CC-pVDZ\cite{dunning_gaussian_1989} basis.  The LCCSD and O2D2-UCCSD curves are essentially overlapping.  CCSDT apparently converged to an excited state at 90$^{\circ}$ so the point was omitted.}
\label{c2h4}
\end{figure}
\end{center}
The divergence of LCCSD and O2D2-UCCSD in figures \ref{hf}-\ref{c2h4} demonstrates the failure of the linear methods to break single bonds.  In general, some excitation-specific correction of higher order in $\mathbf{t}$ is required to treat this type of problem.  For example, ACPF also fails to dissociate these molecules, despite having diagonal corrections of its own, since it still corrects every excitation uniformly.\cite{malrieu_ability_2010}

For HF dissociation (Fig. \ref{hf}), the O2D3-UCCSD approach gives similar qualitative behavior to CCSD(T) if canonical orbitals are used.  Canonical-orbital O2D$\infty$-UCCSD diverges for HF.  We explain the worse performance of the infinite-order correction here by the strong orbital dependence of diagonal methods.  When Kohn-Sham orbitals are used, the unphysical ``hump'' is eliminated from O2D3, and O2D$\infty$ becomes quite accurate for this dissociation.

The CH$_4$ dissociation is largely similar to that of HF, with two notable exceptions.  First, O2D$\infty$ does not completely diverge with canonical orbitals.  Second, the KS orbitals fail to eliminate the unphysical ``hump'' at 2.5\AA{} entirely, with O2D3-UCCSD giving quantitatively better energies at dissociation.  We suspect that this would not be the case with better orbitals. 

In the case of ethylene torsion (Fig. \ref{c2h4}), LCCSD and O2D2-UCCSD's unphysical divergences are never fully corrected, though the O2D3- and O2D$\infty$-UCCSD methods give a clear improvement.  The KS orbitals appear considerably less helpful for this system, introducing very little difference to the diagonally corrected methods beyond easier numerical convergence of the algorithm.  The difference between Trotterized and un-Trotterized methods is small for all three systems, as expected based on the full-order size-extensivity of UCCSD. 

The inconsistent utility of DFT orbitals motivates investigation into optimal orbital choice for the O2D3- and O2D$\infty$-UCCSD methods.  The matter of orbital ``optimization'' is complicated by the fact that divergence to $-\infty$ is possible for some orbital choices, as seen in figures \ref{hf} and \ref{c2h4}.  Minimizing the norm of $\mathbf{t}$ or some similar scheme might be appropriate, but we defer such questions to future work.  \textcolor{black}{This idea that using orbital rotations  to avoid large t-amplitudes is a viable strategy is somewhat corroborated by the relative performance of canonical and Kohn-Sham orbitals.  For example, in the case of hydrogen fluoride dissociation, O2D$\infty$-UCCSD diverges with canonical orbitals, but not with Kohn-Sham orbitals.  This implies that for the Kohn-Sham orbitals, the optimal amplitudes are finite, which is not the case with canonical orbitals.}

The failure of our methods to fully deal with ethylene torsion suggests that we do lose some of the applicability of CEPA(3) as a single-reference method. \cite{malrieu_ability_2010}  The primary difference between ethylene torsion and our single bond-breaking tests is that in ethylene, both the 2-HOMO and 2-LUMO begin to become degenerate as well as the HOMO and LUMO.  (As Malrieu et. al. point out,\cite{malrieu_ability_2010} this is because it is a $\pi/\pi^*$ orbital pair becoming degenerate.)  While our methods include high order interactions between individual excited states and the reference, they neglect high-order coupling between different excited states.  Consequently, we expect our method to break down in situations where there are multiple coupled, excited determinants which are important. 

The O2-UCCSD and O2-tUCCSD methods are extremely similar in their performance.  This can be explained by the fact that UCCSD is fully connected at full order \cite{szalay_alternative_1995} so that the affected terms occur at third-order or higher in the  Taylor series expansion.  A similar argument has been used previously to justify the manual deletion of similar types of ``internally disconnected'' terms from perturbatively truncated CC functional approaches.  \cite{szalay_alternative_1995}  Furthermore, in most situations, O2-UCCSD and O2-tUCCSD differ very little from LCCSD.  All three methods involve solving similar sets of linear systems of equations, and have similar pathologies involving singularities in those equations.\cite{taube_rethinking_2009}

As a broader test of applicability, we considered the CRE-31 reaction energy test set of Soyda{\c s} and Bozkaya,\cite{soydas_assessment_2014} motivated by its use for characterizing orbital-optimized LCCD.  (This test set is enumerated in Table \ref{tab:reactions}.) We elected to use the CC-pVTZ\cite{dunning_gaussian_1989, woon_gaussian_1994} basis for all systems, with geometries obtained from B3LYP/6-31G$^{**}$ optimization in PySCF.
\begin{table}[h]
\begin{tabular}{cr}
 1)& $F_2O+H_2\rightarrow F_2 + H_2O$\\ 
 2)& $H_2O_2+H_2\rightarrow 2H_2O$ \\ 
 3)& $CO + H_2\rightarrow CH_2O$\\
 4)& $CO + 3H_2\rightarrow CH_4 + H_2O$ \\ 
 5)& $N_2+3H_2\rightarrow 2NH_3$\\ 
 6)& $N_2O+H_2\rightarrow N_2 + H_2O$\\ 
 7)& $HNO_2+3H_2\rightarrow 2H_2O+NH_3$\\ 
 8)& $C_2H_2+H_2\rightarrow C_2H_4$\\
 9)& $CH_2CO+2H_2\rightarrow CH_2O + CH_4$\\ 
 10)& $BH_3+3HF\rightarrow BF_3 + 3H_2$\\
 11)& $HCOOH\rightarrow CO_2 + H_2$\\ 
 12)& $CO+H_2O\rightarrow CO_2 + H_2$\\
 13)& $C_2H_2+HF\rightarrow CH_2CHF$\\ 
14)& $HCN+H_2O\rightarrow CO + NH3$\\ 
15)& $HCN+H_2O\rightarrow HCONH_2$\\ 
16)& $HCONH_2+H_2O\rightarrow HCOOH + NH_3$\\
17)&$HCN+NH_3\rightarrow N_2 + CH_4$\\
18)&$CO+CH_4\rightarrow CH3CHO$\\
19)&$N_2+F_2\rightarrow\text{trans}-N_2F_2$\\
20)&$N_2 + F_2\rightarrow \text{cis}-N_2F_2$ \\
21)& $2BH_3\rightarrow B_2H_6$\\
22)& $CH_3ONO\rightarrow CH_3NO_2$\\
23)& $CH_2C\rightarrow C_2H_2$\\ 
24)&  $\text{allene}\rightarrow\text{propyne}$\\
25)& $\text{cyclopropene}\rightarrow\text{propyne}$\\
26)& $\text{oxirane}\rightarrow CH_3CHO$\\
27)& $\text{vinyl alcohol}\rightarrow CH_3CHO$\\
28)& $\text{cyclobutene}\rightarrow 1,3-\text{butadiene}$\\
29)& $2NH_3\rightarrow (NH_3)_2$\\
30)&$2H_2O\rightarrow (H_2O)_2$\\
31)&$2HF\rightarrow (HF)_2$ \\ 
\end{tabular}
\caption{Reaction Key for CRE-31}
\label{tab:reactions}
\end{table}
\normalsize
\begin{center}
\begin{figure}
\includegraphics[width = .5\textwidth]{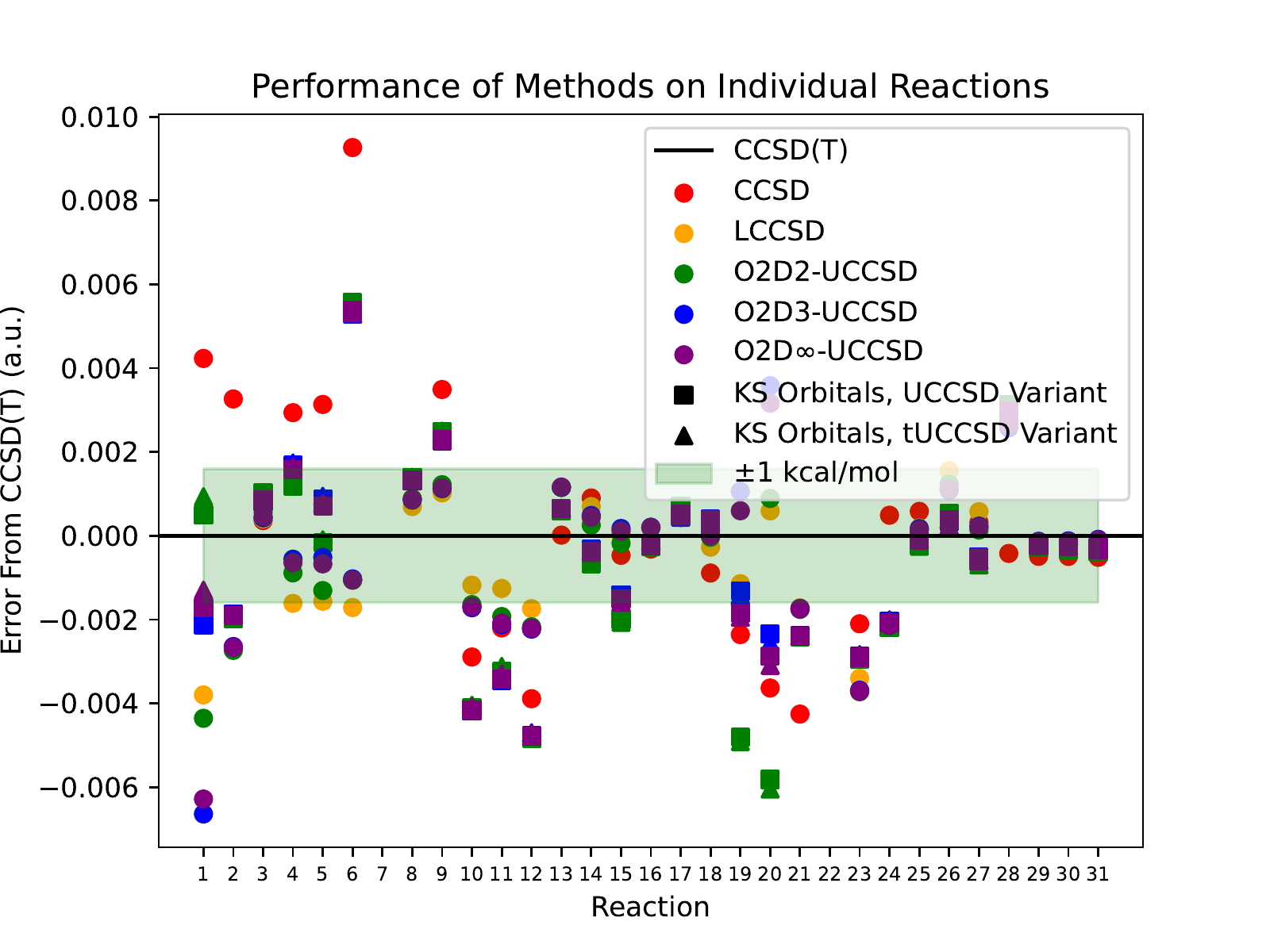}
\caption{Reaction energy errors of the CRE-31 test set in the cc-pVTZ basis, with reactions 7 and 22 excluded due to non-convergence for some methods.}
\label{thing1}
\end{figure}
\begin{figure}
\includegraphics[width = .5\textwidth]{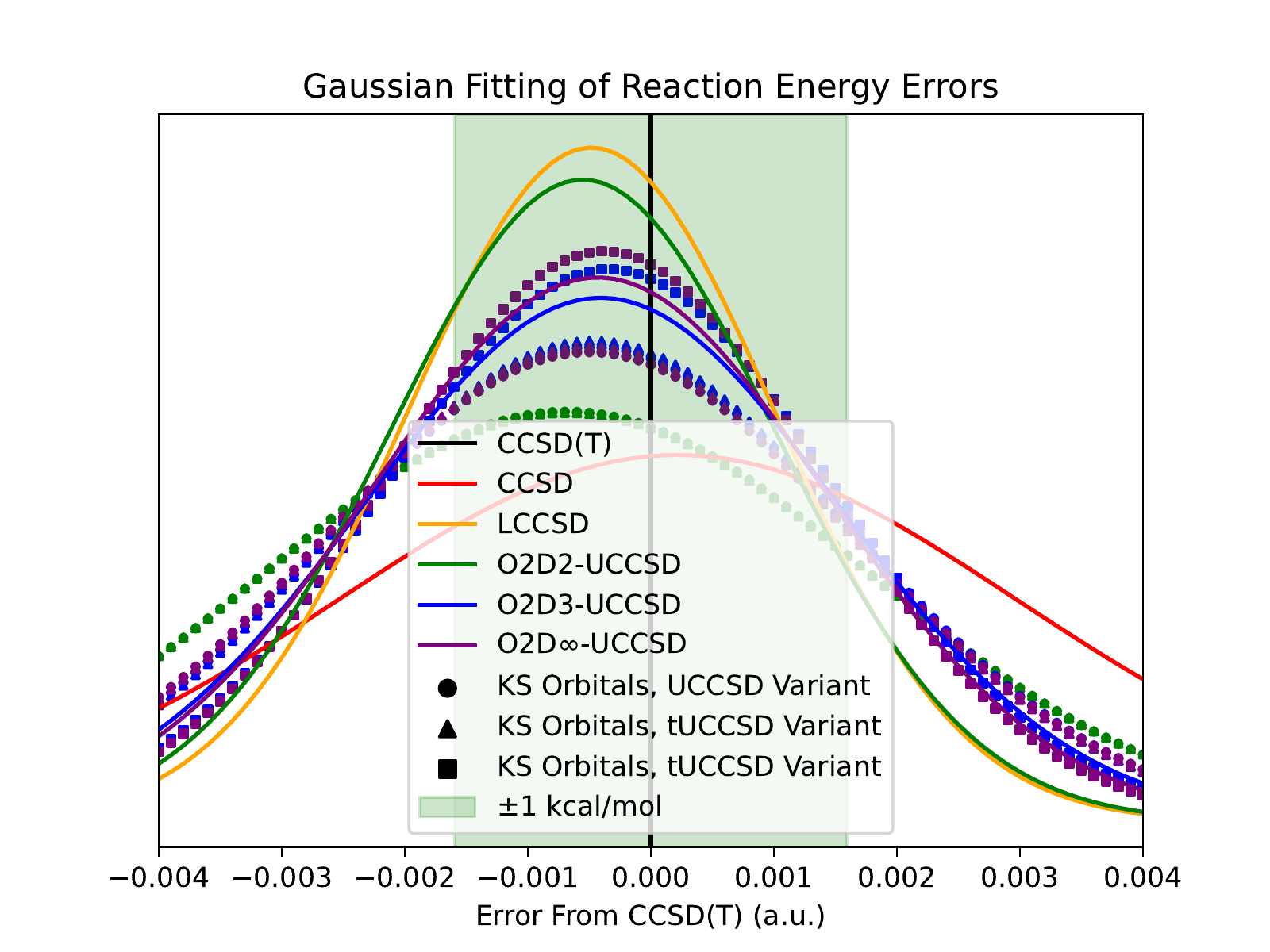}
\caption{Gaussian fittings of the CRE-31 test set in the cc-pVTZ basis, with reactions 7 and 22 excluded due to non-convergence for some methods.}
\label{thing2}
\end{figure}

\end{center}

We begin our analysis of Figs. \ref{thing1} and \ref{thing2} by noting that for some molecules, one or more of our approaches failed to find a local minimum on which to converge.  This is a weakness of using an unconstrained optimizer (L-BFGS-B)\cite{liu_limited_1989} to minimize a functional which is not actually bounded below.  One could imagine a different starting guess allowing O2D3- or O2D$\infty$-UCCSD to converge in these cases, or settling for a minimized gradient as an approximate solution condition.  These three molecules had three of the four highest CCSD $\hat{T}_1$ diagnostics in the test set.  (CH$_3$NO$_2$, CH$_3$ONO, and HNO$_2$ had diagnostics 0.016, 0.018, and 0.018 respectively.)  For context, Lee and Taylor consider a $\hat{T}_1$ diagnostic of .02 indicative of important multireference character.\cite{lee_diagnostic_1989}  We exclude the two associated reactions from the main text.  We note in passing that one can compute a $\hat{T}_1$ diagnostic based on the amplitudes from the O2D2-UCCSD functional, and that this O2D2-UCCSD $\hat{T}_1$ diagnostic is able to predict multireference character that causes the method to catastrophically overestimate the correlation energy.  Individual reaction energies for each method are available in the supplementary information, along with individual molecule error as a function of the O2D2-UCCSD $\hat{T}_1$ diagnostic.

In general, our methods do not appear to be particularly helpful for the CRE-31 test set.  Regardless of orbital choice, Trotterization, or diagonal correction, we achieve performance comparable to LCCSD, and better than CCSD.  We consider this middling performance useful overall - the truncated UCC framework overcomes one of the primary issues of LCCSD, its lack of systematic improvability.  \cite{wennmohs_comparative_2008}  Additionally, the amenability of these methods to multireference implementation should offer a route to avoiding problems with multiple quasi-degeneracies.  
\section*{Conclusions}
We have outlined a pedagogically simple way to correct Taylor-truncated UCC functionals without worsening their formal scaling, using only unmixed derivatives.  While these corrections seem to be of minimal help for computing reaction energies, they dramatically improve the behavior of single-reference O2-UCCSD for single bond-breaking events, repairing one of the most prominent pathologies of LCCSD in an extremely simple, physically motivated way.  We believe that further investigation into orbital optimization can only improve our method given its strong orbital dependence, and plan to explore this in future work.

Additionally, we offer an alternative, Trotterized ansatz which eliminates the extensivity-violating diagrams from second-order UCCSD, noting that tUCC possesses all the qualities that made UCCSD attractive to begin with.  Furthermore, the exactness of certain Trotter orderings of tUCCSD\dots{N} has been rigorously proven \textcolor{black}{by Evangelista et. al.}, while UCCSD\dots{N} may not be exact in certain pathological situations.\cite{evangelista_exact_2019}  \textcolor{black}{It is worth noting that a ``doubles-then-singles'' operator ordering roughly corresponding to that used in our tUCCSD approaches was shown by Evangelista et. al. not to be generally capable of representing arbitrary states, even in the case of only two electrons.  A potentially exact (for two electrons) ``singles-then-doubles'' ordering would not give a size-extensive second-order approximation.}
   
\section*{Acknowledgements}
The authors express their gratitude to the Department of Energy for their support.  (Award No.  DE-SC0019199)
H.R.G. thanks the Institute of Critical Technology and Applied Science for their financial support and Ayush Asthana for useful discussions on formal scaling.  The authors also wish to thank the developers of Psi4, PySCF, and MRCC for the use of their software.  The authors acknowledge Advanced Research Computing at Virginia Tech for providing computational resources and technical support that have contributed to the results reported within this paper. URL: https://arc.vt.edu/
\section*{{Supplementary Information}}
\textcolor{black}{Supplementary information includes:}
\begin{itemize}
    \item \textcolor{black}{.zip file of molecular geometries}
    \item \textcolor{black}{.csv file of individual reaction energies for the CRE-31 test set}
    \item \textcolor{black}{plot of individual reaction energy errors as a function of the O2D2-UCCSD $\hat{T}_1$ diagnostic}
    \end{itemize}
\textcolor{black}{This information is available free of charge via the Internet at http://pubs.acs.org}

\bibliography{manuscript_bw}
\appendix
\section{Derivation of Eq. \ref{inf}}\label{inf_der}
Computing the infinite-order correction to O2-UCCSD is somewhat involved.  We will derive a more general form of Eq. \ref{inf} where one has $t$-amplitudes $\textcolor{black}{t_\mu}$ associated with operator $\hat{O}_\mu$:
\begin{equation}
    \hat{O}_\mu = \hat{a}_\mu^\dagger - \hat{a}_\mu
\end{equation}
We assume only that
\begin{align}
    \hat{a}_\mu^\dagger\ket{\phi_0} &= \ket{\mu}\\
    \hat{a}_\mu^\dagger\ket{\mu} &= 0\\
    \hat{a}_\mu\ket{\phi_0} &= 0\\
    \hat{a}_\mu\ket{\mu} &= \ket{\phi_0}
\end{align}
To find the infinite-order, unmixed part of the energy, we need to find (in Einstein notation with respect to $\mu$):
\begin{equation}
\epsilon_{D\infty} = E_0 + \sum_{k = 1}^\infty \frac{1}{k!}\left.\frac{\partial^k\epsilon_{UCC}}{(\partial t_\mu)^k}\right|_\mathbf{0}t_\mu^k \label{full}
\end{equation}
Using the BCH definition, we know that:
\begin{equation}
    \epsilon_{UCC} = E_0 + \sum_{k=1}^\infty \frac{1}{k!}\braket{\phi_0|[\hat{H}_N,\hat{K}]_k|\phi_0}
\end{equation}
where the subscript $k$ denotes the $k^{th}$ nested commutator.  Consequently,
\begin{equation}
    \left.\frac{\partial^k\epsilon_{UCC}}{(\partial t_\mu)^k}\right|_\mathbf{0} = \braket{\phi_0|[\hat{H}_N,\hat{O}_\mu]_k|\phi_0} \label{general}
\end{equation}
(The factor of $1/k!$ is cancelled by the $k!$ terms that arise when differentiating) Eq. \ref{general} can be simplified into two cases, where we are taking an even or odd derivative.  We first consider the case where it is even.  For $k\in \mathbf{N}$:  
\begin{align}
    &\braket{\phi_0|[\hat{H}_N,\hat{O}_\mu]_{2k}|\phi_0} \label{even}\\&= \sum_{j=0}^{2k}{2k\choose j}\braket{\phi_0|\left(\hat{O}_\mu^\dagger\right)^j\hat{H}_N\left(\hat{O}_\mu\right)^{2k-j}|\phi_0}\nonumber 
\end{align}
For terms in this summand where $j$ is even, we will get some multiple of $\braket{\phi_0|\hat{H}_N|\phi_0} = 0$, so we can simplify equation \ref{even} to
\begin{align}
    &\braket{\phi_0|[\hat{H}_N,\hat{O}_\mu]_{2k}|\phi_0} \\&=
    \sum_{j=0}^{k-1}{2k\choose 2j+1}\braket{\phi_0|\left(\hat{O}_\mu^\dagger\right)^{2j+1}\hat{H}_N\left(\hat{O}_\mu\right)^{2k-2j-1}|\phi_0}\nonumber\\
    &=\braket{\mu|\hat{H}_N|\mu}(-1)^{k-1}\sum_{j=0}^{k-1}{2k\choose 2j+1}\\
    &=\braket{\mu|\hat{H}_N|\mu}(-1)^{k-1}2^{2k-1}\\
    &\equiv \left.\frac{\partial^{2k}\epsilon_{UCC}}{(\partial t_\mu)^{2k}}\right|_\mathbf{0},
\end{align}
using the binomial coefficient for expanding the nested commutator. 
We now consider the situation where we are taking an odd derivative.  For $k\geq 0$:
\begin{align}
    &\braket{\phi_0|[\hat{H}_N,\hat{O}_\mu]_{2k+1}|\phi_0} \label{odd}\\&= \sum_{j=0}^{2k+1}{2k+1\choose j}\braket{\phi_0|\left(\hat{O}_\mu^\dagger\right)^j\hat{H}_N\left(\hat{O}_\mu\right)^{2k+1-j}|\phi_0}\nonumber
\end{align}
The terms where $j$ is odd are the complex conjugates of those where $j$ is even.  We will assume real-valued operators and molecular orbitals, so we can simplify equation \ref{odd} to:
\begin{align}
    &\braket{\phi_0|[\hat{H}_N,\hat{O}_\mu]_{2k+1}|\phi_0} \\&=
    2\sum_{j=0}^{k}{2k+1\choose 2j}\braket{\phi_0|\left(\hat{O}_\mu^\dagger\right)^{2j}\hat{H}_N\left(\hat{O}_\mu\right)^{2k+1-2j}|\phi_0}\nonumber\\
    &=2\braket{\phi_0|\hat{H}_N|\mu}(-1)^{k}\sum_{j=0}^{k}{2k+1\choose 2j}\\
    &=\braket{\phi_0|\hat{H}_N|\mu}(-1)^{k}2^{2k+1}\\
    &\equiv \left.\frac{\partial^{2k+1}\epsilon_{UCC}}{(\partial t_\mu)^{2k+1}}\right|_\mathbf{0}
\end{align}
Summing over all even terms in Eq. \ref{full} gives:
\begin{align}
&\braket{\mu|\hat{H}_N|\mu}\sum_{k=1}^{\infty}\frac{1}{(2k)!}t_\mu^{2k}(-1)^{k-1}2^{2k-1}\\
&=\frac{1}{2}\braket{\mu|\hat{H}_N|\mu}\left(\frac{(2t_\mu)^2}{2!}-\frac{(2t_\mu)^4}{4!}+\dots\right)\\
&=\frac{1}{2}\braket{\mu|\hat{H}_N|\mu}(1-\cos(2t_\mu))\\
&=\sin^2(t_\mu)\braket{\mu|\hat{H}_N|\mu} \label{evens}
\end{align}
Summing over all odd terms in Eq. \ref{full} gives:
\begin{align}
    &\braket{\phi_0|\hat{H}_N|\mu}\sum_{k=0}^\infty\frac{1}{(2k+1)!}(2t_\mu)^{2k+1}(-1)^k\\
    &=\braket{\phi_0|\hat{H}_N|\mu}\left(2t_\mu - \frac{1}{3!}(2t_\mu)^3 + \dots\right)\\
    &=\sin(2t_\mu)\braket{\phi_0|\hat{H}_N|\mu} \label{odds}
\end{align}
Combining $E_0$ with lines \ref{evens} and \ref{odds} gives us $\epsilon_{D\infty}$.  However, we still want to include the mixed second derivatives in O2D$\infty$-UCC.  Consequently, we add in the term:
\begin{equation}
    \frac{1}{2}\braket{\phi_0|[[\hat{H}_N,\hat{K}],\hat{K}]|\phi_0}-t_\mu^2\braket{\mu|\hat{H}_N|\mu}
\end{equation}
for a total O2D$\infty$ energy of: 
\begin{align}
    \epsilon &= E_0 + \frac{1}{2}\braket{\phi_0|[[\hat{H}_N,\hat{K}],\hat{K}]|\phi_0} \\&+ \sum_\mu\sin(2t_\mu)\braket{\phi_0|\hat{H}_N|\mu} \nonumber\\&+ \sum_\mu\left(\sin^2(t_\mu) -t_\mu^2\right)\braket{\phi_\mu|\hat{H}_N|\phi_\mu}\nonumber
\end{align}
Restricting $\mu$ to the singles and doubles yields equation \ref{inf}.
\section{A Minimal EPV CEPA}\label{cepa}
We ignore the singles and triples for simplicity in making our point.  Consider the doubles equations from CIDQ\dots{N}, where we use $\hat{T}$ for consistency with our earlier definitions:
\begin{align}
\label{amps} E_c t_{ab}^{ij} &= \bra{\phi_{ij}^{ab}}\hat{H}_N\left(1 + \hat{T}_2 + \hat{T}_4\right)\ket{\phi_0}\\
\label{ec} E_c &= \sum_{\substack{i<j\\a<b}}\braket{ij||ab}t_{ab}^{ij}
\end{align}
Equations \ref{amps} and \ref{ec} are simply a statement of the CIDQ\dots{N} eigenvalue problem with intermediate normalization.  We can simplify Eq. \ref{amps} by using a CC-type approximation:
\begin{equation}\label{cc}
\hat{T}_4 \approx \frac{1}{2}\hat{T}_2^2
\end{equation}
This leaves us with the quadratic CI term:
\begin{equation}\label{fullcc}
\frac{1}{2}\braket{\phi_{ij}^{ab}|\hat{H}_N\hat{T}_2^2|\phi_0}
\end{equation}\\
Further approximating the quadruples in this equation by:
\begin{equation}
\hat{T}_2^2\ket{\phi_0} \approx \{\hat{T}_2,\hat{a}_{ij}^{ab}t_{ab}^{ij}\}\ket{\phi_0} = 2\hat{T}_2 \hat{a}_{ij}^{ab}t_{ab}^{ij}\ket{\phi_0}
\end{equation}
lets us simplify term \ref{fullcc} to 
\begin{align}
\braket{\phi_{ij}^{ab}|\hat{H}_N\hat{T}_2|\phi_{ij}^{ab}}t_{ab}^{ij} = E_ct_{ab}^{ij} - \sum_{\substack{k<l\\c<d}\cup ijab}\braket{kl||cd}t_{cd}^{kl}t_{ab}^{ij}
\end{align}
The $\cup$ summation is over the exclusion principle-violating (EPV) terms where c, d, k, or l is equivalent to a, b, i, or j.  Neglecting these terms entirely gives the LCCD equations:
\begin{align}
0 &= \braket{\phi_{ij}^{ab}|\hat{H}_N\left(1 + \hat{T}_2\right)|\phi_0}\\
E_c &= \sum_{\substack{i<j\\a<b}}\braket{ij||ab}t_{ab}^{ij}
\end{align}
Something similar to our method emerges if one instead makes the approximation that
\begin{equation}
\sum_{\substack{k<l\\c<d}\cup ijab}\braket{kl||cd}t_{cd}^{kl}t_{ab}^{ij} \approx \braket{ij||ab}\left(t_{ab}^{ij}\right)^2
\end{equation}
That is, we only care about one EPV term, where every single index is EPV.  This gives amplitude equation:
\begin{equation}
-\braket{ij||ab} = \braket{\phi_{ij}^{ab}|\hat{H}_N\hat{T}_2|\phi_0} - \braket{ij||ab}\left(t_{ab}^{ij}\right)^2
\end{equation}
Including a factor of 2 in front of $\braket{ij||ab}\left(t_{ab}^{ij}\right)^2$ would give the stationary condition of the O2D3-UCCD method.  This suggests that our diagonal corrections are similar in spirit to a simplified CEPA(3), where all EPV terms with k or l equal to i or j are considered. 
\section*{For Table of Contents Only}
\includegraphics{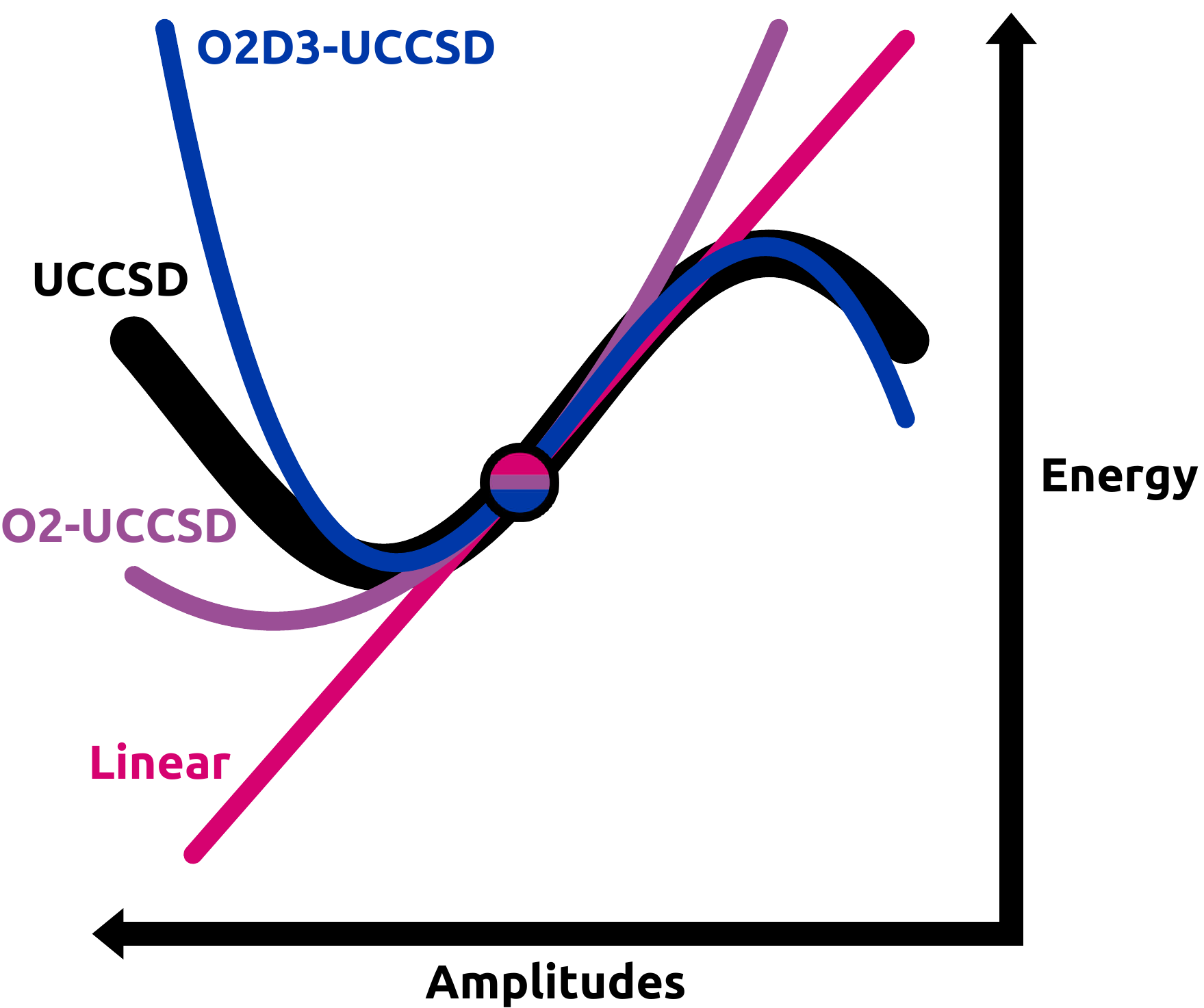}
\end{document}